\documentclass[prd,showpacs,amssymb,amsmath,preprint]{revtex4}
\usepackage{graphicx}

\begin{document}
\title{Decoherence induced by long wavelength gravitons}
\author{V. A. \surname{De Lorenci}}
 \email{delorenci@unifei.edu.br}
\author{L. H. \surname{Ford}}
 \email{ford@cosmos.phy.tufts.edu}
\affiliation{$\mbox{}^{*}$Instituto de F\'{\i}sica e Qu\'{\i}mica,
Universidade Federal de Itajub\'a, Itajub\'a, MG 37500-903, Brazil} 
\affiliation{$\mbox{}^\dag$Institute of Cosmology, Department of Physics and Astronomy,
Tufts University, Medford, MA 02155, USA}

\begin{abstract}
We discuss how a background bath of gravitons can induce decoherence of quantum systems. The
mechanism is dephasing, the loss of phase coherence due to quantum geometry fluctuations caused
by the gravitons. This effect is illustrated in a simple analog model of quantum particles in a cavity
whose walls undergo position fluctuations, and create the same effect expected from spacetime
geometry fluctuations. We obtain an explicit result for the decoherence rate in the limit where the
graviton wavelength is large compared to the size of the quantum system, and make some estimates
for this rate.
\end{abstract}


\maketitle

\baselineskip=14pt

\section{Introduction}

The interaction of the environment with a quantum system tends to lead to a loss of 
quantum coherence (decoherence), and a transition from quantum to classical behavior.
Decoherence has been the topic of extensive investigation in recent  years. 
(For a review, see Ref.~\cite{schlosshauer2007}.) Several authors~\cite{Diosi,Penrose,Kay}  
have suggested that
the gravitational interaction might play a special role in the quantum to classical transition,
but this view remains controversial. For a recent review with further references, see Hu~\cite{Hu}.
However, there seems to be no doubt that gravitational interactions can contribute to
decoherence.

In this paper, we will treat  systems where the role of gravity is to provide spacetime geometry
fluctuations which in turn lead to length and relative phase fluctuations. This effect can
arise from the presence of a bath of long wavelength gravitons. First consider gravity waves
on a flat background described by a metric of the form (in $c=1$ units)
\begin{equation}
ds^2 =-dt^2 + h_{ij} \, dx^i dx^j\, ,
\label{eq:metric}
\end{equation}
where the transverse tracefree gauge is assumed. Fluctuations of the metric lead to fluctuations
of the squared proper length $d \ell^2 = h_{ij} \, dx^i dx^j$, and hence of the separation between nearby
geodesics. This follows from the geodesic deviation equation for the  separation $\xi^i$, 
\begin{equation}
\frac{d^2 \xi_i}{d t^2} = - R_{itjt} \, \xi^j\,,
\label{eq:geo-dev}
\end{equation}
where the relevant component of the Riemann tensor is given by $R_{itjt} = -\frac{1}{2} \partial_t^2 h_{ij}$
in the transverse tracefree gauge.  In many situations, length fluctuations can lead to quantum phase
fluctuations, and hence decoherence through dephasing. In Sec.~\ref{sec:model}, we will treat a
simple model in which the length fluctuations arise from fluctuating boundaries, which forms an
analog model for the effects of quantum geometry fluctuations. We apply the lessons of this model
to the effects of a graviton bath in Sec.~\ref{sec:bath}, and obtain estimates for the decoherence rate.
Our results are summarized in Sec.~\ref{sec:sum}.

\section{A Model with Fluctuating Boundaries}
\label{sec:model}

\begin{figure}[!hbt]
\leavevmode
\centering
\includegraphics[scale = 1.0]{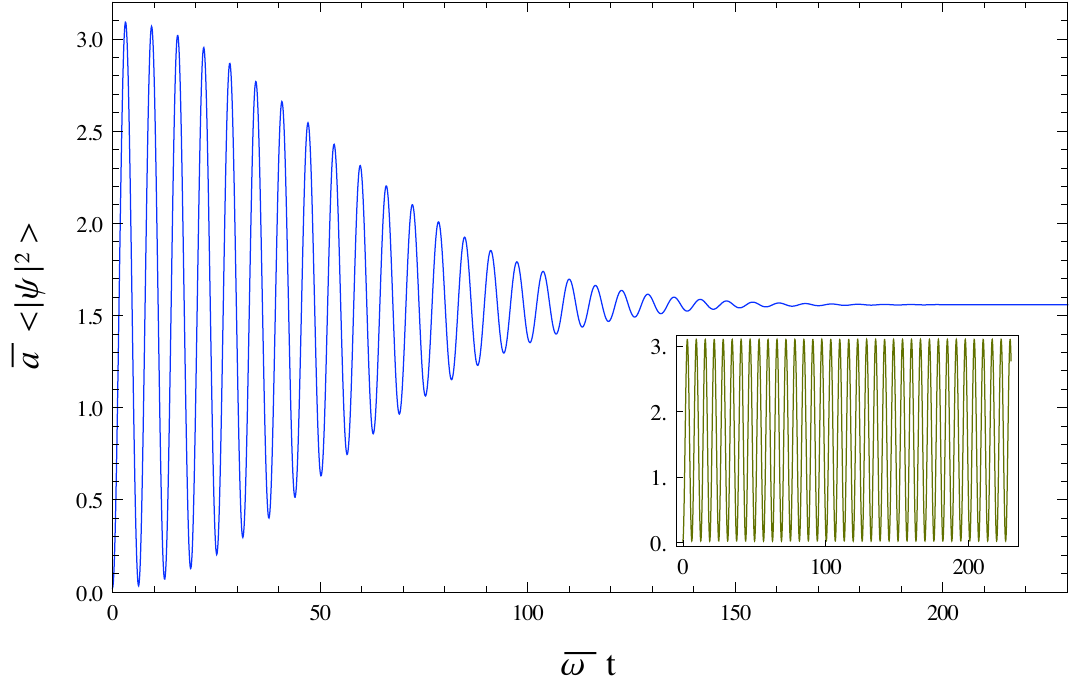}
\caption{{\small\sf (color online). 
The figure shows the behavior of the probability density associated with 
the  state defined by Eq. (\ref{5}) when fluctuating boundaries 
are assumed. As time goes on, interference effects between 
the component states of $\psi(x,t)$ are suppressed.  
For fixed boundaries, no suppression of interference 
is found, as shown in the down inset frame. 
We set $x/\bar{a}=0.7$ and $\sigma=0.01$.}}
\label{fig1}
\end{figure}

We begin with a non-relativistic particle of mass $m$ confined in an 
infinite potential well whose width is denoted by $a$. For the sake of simplicity, we
restrict our analysis to the case of one space dimension. 
Suppose that the normalized state of the particle at an arbitrary
time $t$ is given by a superposition of the first two available states as
\begin{equation}
\psi(x,t) = \frac{1}{\sqrt{2}}\left[\psi_1(x,t) + \psi_2(x,t)\right].
\label{5}
\end{equation}
The eigenfunctions $\psi_n$ ($n=1,2$) are independent solutions of the 
Schr\"odinger equation under the conditions that these functions 
vanish on the boundaries at $x=0$ and $x=a$, yielding 
\begin{equation}
\psi_n(x,t) = \sqrt{\frac{2}{a}}\,\sin\left(\frac{n \pi x}{a}\right) \;{\rm e}^{-i\omega_n t},
\label{1}
\end{equation}
with $\omega_n = n^2 \pi^2 \hbar / 2 m a^2$. Each one of these eigenfunctions
describes a stationary state, as the corresponding probability density $|\psi_n|^2$ 
is time-independent. However, time evolution occurs when the
particle is governed by a linear superposition of $\psi_n(x,t)$, such as that 
given in Eq. (\ref{5}). The probability density can be obtained as
\begin{eqnarray}
|\psi(x,t)|^2 = \frac{1}{2}\left(|\psi_1|^2 + |\psi_2|^2\right) 
+ |\psi_1||\psi_2|\cos(\Delta\omega \, t) \,,  
\label{11}
\end{eqnarray}
where we define $\Delta\omega = \omega_2 - \omega_1$, the Bohr angular frequency 
associated with the energy difference of the superposed states.
As we see, the time evolution of $|\psi(x,t)|^2$ is exclusively governed 
by the interference term between $\psi_1$ and $\psi_2$. In fact, $|\psi(x,t)|^2$
oscillates between its maximum $(|\psi_1|+|\psi_2|)^2/2$ and minimum
$(|\psi_1|-|\psi_2|)^2/2$ values, as depicted in the down inset frame in 
Fig. \ref{fig1}, for a particular numerical model. 

Now we wish to investigate the behavior of this quantum system when interaction
with the environment takes place. In order to model the interaction
between the system and its environment, we allow the positions of the
physical boundaries to fluctuate under the influence of an external
noise. This can simply be implemented by allowing the width 
parameter $a$ to undergo fluctuations around a mean value $\bar a$ \cite{ford1998}. 
Note that fluctuations in the width, $a$, lead to fluctuations in the energy
levels, $\omega_n$. 

An accelerating boundary can emit quantum radiation which might interact with
the particle~\cite{FD76,DF77,FV82}. However, quantum fluctuations of position
need not imply classical acceleration or radiation. We view the boundary as being 
analogous to an electron in a quantum state, such as a Gaussian wavepacket,
which is not an eigenstate of position, but yet need not radiate.

We set $a = \bar{a}(1+\varepsilon)$, where the dimensionless parameter $\varepsilon$
is described by a Gaussian distribution as
\begin{equation}
f(\varepsilon) = \sqrt{\frac{\theta}{\pi}}\;{\rm e}^{-\theta \varepsilon^2},
\label{13}
\end{equation}
with $\theta$ related to the width $\sigma$ of the distribution by means
of $\sigma^2 = 1/(2\theta)$. The mean value of an arbitrary function $G(\varepsilon)$
over $\varepsilon$ is a linear operation defined by
\begin{equation}
\left<G\right>  = \int_{-\infty}^{\infty}G(\varepsilon)f(\varepsilon)d\varepsilon.
\label{15}
\end{equation}
Notice that  $\left<\varepsilon^2\right> -\left<\varepsilon\right>^2 = \sigma^2$, 
the mean squared fluctuation of $\varepsilon$.

There does not seem to be a meaning to averaging the wave function $\psi(x,t)$, as it is 
not directly observable. 
(See, however, Ref.~\cite{lundeen2011} for a discussion of the possibility of measuring a wave 
function in the context of a weak measurements approach.) 
Here we are interested in studying averaged values of observable quantities. Particularly,
the modulus squared of the particle wave function, Eq.~(\ref{11}), represents the probability 
density associated with the  position the particle .  
The average over width fluctuations of this quantity can be calculated by using
Eqs. (\ref{11}) and (\ref{15}), yielding
\begin{eqnarray}
\left<|\psi|^2\right> = \frac{1}{2}\left(\left<|\psi_1|^2\right> + \left<|\psi_2|^2\right>\right) 
+ \left<|\psi_1||\psi_2|\cos(\Delta\omega \,t\right>.
\label{22}
\end{eqnarray}
Here the angular bracket refers to the average over positions of the boundaries, which
was defined in Eq.~(\ref{15}).
In what follows, each term appearing in the above equation will be considered separately.

The two first terms appearing in Eq.~(\ref{22}) can be  expressed, using Eqs.~(\ref{1}) and 
(\ref{15}), as  
\begin{equation}
\left<|\psi_n|^2\right> =  \sqrt{\frac{4\theta}{\pi}} \int_{-\infty}^{\infty} 
\frac{{\rm e}^{-\theta \varepsilon^2}}{\bar{a}(1+\varepsilon)} \sin^2 \left[\frac{n\pi x}
{\bar{a}(1+\varepsilon)} \right] \, d\varepsilon \,,
\label{23}
\end{equation}
where $n = 1,2$.
We assume a narrow distribution, $\sigma \ll 1$, which means that only small values
of $\varepsilon$ contribute in Eq.~(\ref{23}). 
We  expand the integrand, apart from the exponential, to second order in $\varepsilon$,
and then integrate to obtain
\begin{eqnarray}
\left<|\psi_n|^2\right> &=&
\frac{2}{\bar a}\sin^2\left(\frac{n \pi x}{\bar a}\right) + 
\left[\frac{2}{\bar a}\sin^2\left(\frac{n \pi x}{\bar a}\right) + 
\frac{4n\pi x}{\bar{a}^2}\sin\left(\frac{2n \pi x}{\bar a}\right)\right.
\nonumber \\
&&+\left. \frac{2n^2\pi^2 x^2}{\bar{a}^3}\cos\left(\frac{2n \pi x}{\bar a}\right) 
\right]\sigma^2 + O(\sigma^4).
\label{31}
\end{eqnarray}
Further, as the particle is confined ($0\le x \le \bar a$), and the small $\sigma$
 approximation, $\pi\sigma x/\bar a \ll 1$, is assumed, we have
\begin{equation}
\left<|\psi_n|^2\right>\, \approx 
\frac{2}{\bar a}\sin^2\left(\frac{n \pi x}{\bar a}\right) + O(\sigma^2) \approx  |\psi_n|^2.
\label{34}
\end{equation}
In lowest order in $\sigma$, the probability density associated with the energy eigenstates
 does not   change  when
boundary fluctuations are introduced.  As may be seen from Eq.~(\ref{31}), 
there is a shift in this density in order $\sigma^2$. However, this shift is
time-independent, and not of special interest here. If the particle is initially
in an energy eigenstate state $\psi_n(x,t)$, it will remain in this state and its
probability density will not undergo appreciable time evolution
when small boundary fluctuations are present.

Next, we consider the last term in Eq.~(\ref{22}), which describes the interference
effects occurring in the system. From Eqs.~(\ref{1}) 
and (\ref{15}), we obtain that 
\begin{eqnarray}
\left<|\psi_1||\psi_2|\cos\omega t\right> &=& 
\frac{1}{4\bar a}\sqrt{\frac{\theta}{\pi}}\left(A_{1}^{+}
+A_{-1}^{+}-A_{3}^{+}-A_{-3}^{+} 
\right. \nonumber\\ 
&& \left.+A_{1}^{-}+A_{-1}^{-}-A_{3}^{-}-A_{-3}^{-}\right), 
\label{35} 
\end{eqnarray}
where $A_q^{\pm}$ is defined by
\begin{eqnarray}
A_q^{\pm}=  \int_{-\infty}^{\infty} \frac{1}{1+\varepsilon}
\exp\left[\frac{q i \pi x}{\bar a (1+\varepsilon)} \pm 
\frac{i{\Delta \omega} t}{(1+\varepsilon)^2} - \theta\varepsilon^2\right]\, d\varepsilon ,
\label{36}
\end{eqnarray}
and $\Delta \omega  = \omega_{2}(\bar a) - \omega_{1}(\bar a) = 3\pi^2/2m\bar{a}^2$
is the energy difference of the two levels.
Once the small $\sigma$ approximation is assumed, in order to solve the above integral
it is enough to Taylor expand to first order in $\varepsilon$ inside the exponential, 
but only to zeroth order otherwise. Thus, using the identity
\begin{equation}
\int_{-\infty}^{\infty} e^{\pm iZ\varepsilon -\theta\varepsilon^2} d\varepsilon
= \sqrt{\frac{\pi}{\theta}} e^{-Z^2/4\theta} \,.
\label{29}
\end{equation} 
and neglecting terms which become unimportant after a finite time 
$t \agt 1/\Delta \omega$, it follows that
\begin{eqnarray}
A_q^{\pm}\approx \sqrt{\frac{\pi}{\theta}}
\exp\left[\frac{iq\pi}{\bar a}
\left(x\pm\frac{\bar a \Delta\omega t}{q\pi}\right)\right]
e^{-\Gamma t^2},
\label{43}
\end{eqnarray}
where $\Gamma = 2\Delta \omega^2\sigma^2$. 
Using this result in Eq.~(\ref{35}), we obtain
\begin{equation}
\left<|\psi_1||\psi_2|\cos\omega t\right> = 
\frac{4}{\bar a}\cos\left(\frac{\pi x}{\bar a}\right)\sin^2\left(\frac{\pi x}{\bar a}\right) 
\cos(\Delta \omega t) \;  e^{-\Gamma t^2}.
\label{44} 
\end{equation}
As one can see this term describes oscillations modulated by a factor which decays
exponentially in squared time.
 The time scale for the onset of this  decay is 
 \begin{equation}
 t_o = \frac{1}{\Delta \omega}\;.
 \end{equation}
 In the
case of an electron in a potential well with $\bar a \approx 1 \mbox{\AA}$, this time is 
of order  $t_o \approx10^{-17}\, s$. However, once the decay begins, the 
characteristic decay time is
\begin{equation}
 t_d = \frac{1}{\sqrt{\Gamma}} = \frac{t_o}{\sqrt{2}\, \sigma}\,,
 \label{eq:decay_time}
 \end{equation}
which is longer by a factor of about $1/\sigma$.

Combining  the results in Eqs.~(\ref{34}) and (\ref{44}) with Eq.~(\ref{22}), we
find that the average $\left<|\psi(x,t)|^2\right>$ becomes
\begin{equation}
\left<|\psi|^2\right> \approx \frac{1}{2}\left(|\psi_1|^2 + |\psi_2|^2\right)
+\frac{4}{\bar a}\cos\left(\frac{\pi x}{\bar a}\right)\sin^2\left(\frac{\pi x}{\bar a}\right) 
\cos(\Delta \omega t) e^{-\Gamma t^2}.
\label{45}
\end{equation}
As time passes, the last term in the above equation  falls to zero for $t \gg t_d$. 
Thus, the net effect of the fluctuations is to kill the interference
term. Neglecting the last term in Eq. (\ref{45}) we obtain that 
\begin{eqnarray}
\left<|\psi|^2\right> = \frac{1}{2}\left(|\psi_1|^2 + |\psi_2|^2\right), 
\label{46}
\end{eqnarray}
which corresponds to a weighted sum of probabilities as occurs when a 
statistical mixture of states is considered. 

The behavior described by Eq.~(\ref{45}) can be illustrated by a numerical example. 
For instance, let us study the time evolution of the averaged probability 
density when fluctuating boundaries are considered in a particular model
with $x/\bar a = 0.7$ and $\sigma = 0.01$. In this case Eq.~(\ref{22}) 
can be integrated numerically. The result is depicted in Fig.~\ref{fig1}. 
Alternatively we could have used the approximate solution given by Eq.~(\ref{45}),
which leads to an identical graph, confirming the approximation used in
obtaining Eq.~(\ref{45}).   
As the figure shows clearly, the probability density $\left<|\psi|^2\right>$ oscillates
around the mean value $(\left<|\psi_1|^2\right> + \left<|\psi_2|^2\right>)/2$,
converging to this value when $t \gg t_d$. As anticipated, the net effect of 
 the fluctuating boundaries of the potential well is to 
kill the interference effect between the two state components $\psi_1$ and $\psi_2$.  
Finally, when no fluctuations are present, the usual stationary solution holds, as
illustrated by the down inset frame in Fig. \ref{fig1}.

In order to have an estimate, consider again the case of an electron in a potential
well with $\bar a\sim 1\mbox{\AA}$. As shown in Fig.~\ref{fig1} the oscillations 
in $\left<|\psi|^2\right>$
are completely suppressed when $\bar\omega t = 200$, that is, $10^{-14}$ seconds after
the boundary fluctuations are turned on.

\section{Spacetime geometry fluctuations from a bath of gravitons}
\label{sec:bath}

Now we wish to turn to quantum systems where the length fluctuations are due to spacetime 
geometry fluctuations. However, we first review the effects of a classical gravity wave, described
by the metric in Eq.~\eqref{eq:metric}, which can cause variations in the positions of a boundary.
The right hand side of Eq.~\eqref{eq:geo-dev} is the tidal acceleration,
or force per unit mass on the walls of the cavity due to gravity. Non-gravitational forces will
modify this equation with the addition of other terms. Consider the case where the walls are 
bound in a harmonic potential with natural frequency $\omega_0$. If the walls are displaced
from their equilibrium position by a distance $\delta \xi$, then the restoring force per unit mass
is $-\omega_0^2 \, \delta \xi$. In the transverse, tracefree gauge, the Riemann tensor for 
the spacetime described by Eq.~\eqref{eq:metric} may be expressed as
\begin{equation}
R_{itjt} = -\frac{1}{2}\, \ddot{h}_{ij} \,.
\end{equation}
Let the displacements be in the $x$-direction, and write $ \xi^i = \delta^i_x\,  \xi$. Now
Eq.~\eqref{eq:geo-dev} is modified to
\begin{equation}
\ddot{\xi} = \ddot{\delta \xi} =  \frac{1}{2}\, \ddot{h}_{xx} \, \xi  - \omega_0^2 \, \delta \xi \,.
\end{equation}
This is the equation for a driven harmonic oscillator, with damping effects neglected.
Let $\omega$ be the frequency of the gravity wave and hence also of the response $\delta \xi$, 
so $\ddot{h}_{xx} =- \omega^2\, {h}_{xx}$, and $\ddot{\delta \xi} = - \omega^2\, \delta \xi$.
This leads to a result for the fractional change in position of the wall:
\begin{equation}
\frac{\delta \xi}{\xi} = \frac{1}{2}\, {h}_{xx} \, \left(\frac{\omega^2}{\omega^2 - \omega_0^2} \right)\,.
\label{eq:fractional} 
\end{equation}
Because damping effects have been ignored, this result is not expected to hold near 
resonance, $\omega \approx \omega_0$, but can be a good approximation well away from
resonance. In particular, if the gravity wave frequency is above the natural frequency of the bound
system, $\omega \gg \omega_0$, then the magnitude of the fractional change in position is of the 
same order as the metric perturbation
\begin{equation}
\left| \frac{\delta \xi}{\xi} \right| \approx \frac{1}{2}\, {h}_{xx}\,.
\label{eq:above}
\end{equation}
Note that this includes walls moving on geodesics as the special case where $\omega_0 =0$.
If the gravity wave frequency is below resonance, $\omega \ll \omega_0$, then $|{\delta \xi}/{\xi}|$
is suppressed by a factor of $(\omega/\omega_0)^2$, so we will focus on the former case described 
by  Eq.~\eqref{eq:above}. 

The use of the geodesic deviation equation, Eq.~\eqref{eq:geo-dev}, to describe the relative motion
of the components of a quantum system, such as the boundaries of a cavity, assumes that the wavelength
of the gravity wave is larger than the geometric size of the system. This arises because we are assuming
that the Riemann tensor is approximately constant on a length scale $\xi$ in writing Eq.~\eqref{eq:geo-dev}.
Thus the gravity wave frequency is bounded both from above and from below:
\begin{equation}
\frac{2 \pi}{\xi} > \omega > \omega_0\,.
\label{eq:bound}
\end{equation}

Now we wish to replace the classical gravity wave with a fluctuating spacetime geometry. One way to
do this is with a bath of gravitons. We will use some results obtained in Refs.~\cite{FP77,F95,FS96}.
Suppose that gravitons are in a state where
\begin{equation}
\langle h_{ij} \rangle =0
\end{equation}
but
\begin{equation}
h^2 = \frac{1}{9}\,\langle h_{ij} \, h^{ij}\rangle \not= 0\,.
\label{eq:h2}
\end{equation}
Examples of such a state include  thermal states and squeezed vacuum states. 
If the characteristic frequency of the gravitons satisfies Eq.~\eqref{eq:bound},
the root-mean-square fractional length fluctuations are of order $h$, which now plays the 
role of the parameter $\sigma$ in the previous section. The factor of $1/9$ in Eq.~(\ref{eq:h2})
is motivated by the expectation that in an isotropic bath will all polarization states equally
excited, we will have  $\langle h^2_{xx} \rangle = \langle h^2_{yy} \rangle = \langle h^2_{zz} \rangle
= \langle h^2_{xy} \rangle = \langle h^2_{yx} \rangle = \langle h^2_{xz} \rangle = \langle h^2_{zx} \rangle
= \langle h^2_{yz} \rangle = \langle h^2_{zy} \rangle$, and hence $h^2  = \langle h^2_{xx} \rangle$.
Then $h$ will be the root-mean-square fractional length fluctuations in a given direction, such as
the $x$-direction. [Note that $h^2$ was defined with a different numerical factor in Refs.~\cite{F95,FS96}.]

In a graviton bath, there will be quantum phase fluctuations, just as in the model in 
Sect~\ref{sec:model}, leading to dephasing and a loss of contrast in an interference
pattern. If the probability distribution for the quantum metric fluctuations is approximately
Gaussian, then the contrast will decay as an exponential of the squared time, as in
Eq.~\eqref{44}. This leads to essentially the same decoherence time as in the fluctuating
boundary model
\begin{equation}
t_d \approx \frac{1}{h\, \Delta \omega}\,.
\label{eq:td} 
\end{equation}
As before, $\Delta \omega$ is the energy difference between the interfering states.
Even if the  probability distribution is not Gaussian, Eq.~\eqref{eq:td} is a reasonable
estimate for the decoherence time, although the decay rate may not have the 
functional form of Eq.~\eqref{44}.

Consider the example of a thermal state of gravitons at temperature $T$. In Ref.~\cite{F95},
it was shown that for such a state
\begin{equation}
\langle h_{ij} \, h^{ij}\rangle = \frac{16}{3} \pi \, \ell_P^2 \,T^2 \,,
\end{equation}
where $\ell_P$ is the Planck length. (Note that units in which $32 \pi G = 32 \pi \ell_P^2 =1$,
where $G$ is Newton's constant, were used in Ref.~\cite{F95}.) 
This leads to an estimate of the decoherence rate given by
\begin{equation}
\frac{1}{t_d} \approx   \frac{4}{3} \sqrt{\frac{\pi}{3}}\; T \, \ell_P\, \Delta \omega = 
\frac{8\pi}{3} \sqrt{\frac{\pi}{3}}\; T \, \frac{\Delta \omega}{E_P}\,.
\label{eq:d-rate}
\end{equation}
Here $E_P = 2 \pi/\ell_P$ is the Planck energy, and we are using units with $\hbar =k_B =1$, where $k_B$
is Boltzmann's constant. 

This result may be compared with a formula recently given by Blencowe~\cite{Blencowe} 
for the decoherence rate by a thermal bath of gravitons
\begin{equation}
\left(\frac{1}{t_d}\right)_{\rm Blencowe}  = T \, \left(\frac{\Delta \omega}{E_P}\right)^2  \,.
\label{eq:d-rate-blen}
\end{equation}
This apparent discrepancy probably arises from use of different assumptions. Our result, 
Eq.~\eqref{eq:d-rate}, assumes a low temperature limit in the sense that the graviton wavelengths 
must be long compared to the geometric dimensions of the quantum system. In contrast, 
Blencowe assumes a high temperature limit. 
In a regime where our result, Eq.~\eqref{eq:d-rate} is applicable, it predicts a larger decoherence rate
than does Eq.~\eqref{eq:d-rate-blen}. This arises because gravitons with wavelengths short compared to 
the size of the system can be expected to be less effective in producing dephasing.  
We may write the decoherence time predicted by Eq.~\eqref{eq:d-rate} as
\begin{equation}
t_d \approx 2 \times 10^9 {\rm yr} \, \left(\frac{1 {\rm K}}{T} \right)\, \left(\frac{1 {\rm eV}}{\Delta \omega} \right)\,.
\label{eq:d-rate-est}
\end{equation}
Thus the gravitational decoherence rate is very small unless either the energy difference of the interfering
states, or the effective graviton temperature is large. 

A truly thermal bath of gravitons could be difficult to produce  due to the weak coupling of gravitons to
one another and to matter, leading to very long equilibration times. Hawking radiation from black holes~\cite{Hawking}
is one mechanism to produce thermal gravitons. In addition, quantum particle creation in an expanding
universe can sometimes produce an approximately Planckian spectrum of particles~\cite{Parker},
including gravitons. Quantum creation of gravitons and other particles are expected at the end of an
inflationary era, and could contribute significantly to the matter and radiation in the universe after the end
of inflation~\cite{F87}. In addition, quantum stress tensor fluctuations during inflation might  contribute
to the graviton background~\cite{WHFN11}. 

We can now generalize Eq.~\eqref{eq:d-rate} to more general baths of gravitons. The energy density 
of such a bath may be written as (See Eq.~(50) in Ref.~\cite{FS96}, where $16 \pi G = 16 \pi \ell_P^2 =1$
units were used.)
\begin{equation}
\rho_g = \frac{1}{32 \pi \ell_P^2}\,  \langle \dot{h}_{ij} \, \dot{h}^{ij}\rangle = 
\frac{\omega_g^2}{32 \pi \ell_P^2}\, \langle h_{ij} \, h^{ij}\rangle\,.
\end{equation}
Here $\omega_g$ is the characteristic graviton angular frequency. We can use Eq.~(\ref{eq:h2})
to write
\begin{equation}
h = \frac{4}{3} \sqrt{2\pi} \, \frac{\lambda_g \, \sqrt{\rho_g}}{E_P}\,,
\end{equation}
where $\lambda_g =2\pi/\omega_g$ is the characteristic graviton wavelength. This leads to a
decoherence time of
\begin{equation}
t_d = \frac{3}{4 \sqrt{2\pi}} \, \frac{E_P}{\lambda_g \, \sqrt{\rho_g}\, \Delta \omega}\,.
\label{eq:d-rate-gen}
\end{equation}
Thus a bath of lower frequency gravitons will be more effective in causing decoherence for a fixed graviton
energy density. We can express the decoherence time as
\begin{equation}
t_d \approx 2 \times 10^8 {\rm yr} \, \left(\frac{1\, {\rm cm}}{\lambda_g} \right)\, \left(\frac{\rho_{\rm CMB}}{\rho_g}\right)^{1/2}
\, \left(\frac{1 {\rm eV}}{\Delta \omega} \right)\,,
\label{eq:d-rate-est2}
\end{equation}
where  $\rho_{\rm CMB}$ is the energy density
associated with the cosmic microwave background at $T \approx 2.7 {\rm K}$. This time is still quite long unless
either $\lambda_g$, $\rho_g$,  or the energy difference $\Delta \omega$ are large. 

Recall that this estimate holds for the case where $\omega_g > \omega_0$, the graviton frequency is larger 
than the natural resonant frequency associated with non-gravitational binding forces.  In the opposite limit,
where $\omega_g < \omega_0$, we see from Eq.~(\ref{eq:fractional}) that the magnitude of the position
fluctuations will be suppressed by a factor of  $(\omega_g /\omega_0)^2$. This leads to a corresponding
decrease in the decoherence rate, and an increase in the  decoherence time by a factor of 
$(\omega_0 /\omega_g)^2$.

\section{Summary}
\label{sec:sum}

In this paper, we have treated an example of gravitational decoherence, the loss of quantum coherence
as a result of gravitational effects. The specific effect we discuss is dephasing due to quantum length
fluctuations. This effect was illustrated by an analog model of quantum particles in a box with walls
whose positions fluctuate. Quantum geometry fluctuations produced by a bath of gravitons were discussed
and shown to have the same effect. The decoherence rate is typically rather small in the present day
universe.  However, the rate increases with increasing graviton wavelength for fixed energy density, and
also increases as the energy differences in the quantum system increase. In any case,
this  is a simple model
which shows that gravitational decoherence is in principle possible, and illustrates the role of spacetime
geometry fluctuations.

\begin{acknowledgments}
This work was partially supported by the Brazilian research agencies CAPES, CNPq, and
FAPEMIG and by the National Science Foundation under Grant  PHY-1205764.
\end{acknowledgments}

\end{document}